\documentclass[5p,hidelinks]{elsarticle}

\usepackage{lineno}
\usepackage{hyperref}
\usepackage{amssymb, amsmath,color}
\usepackage{CJK}

\modulolinenumbers[5]

\begin{document}
\begin{CJK}{UTF8}{mj}

\begin{frontmatter}

\title{On the Heterogeneous Distributions in Paper Citations}

\author{Jinhyuk Yun}
\author{Sejung Ahn}
\author{June Young Lee}\cortext[correspondingauthor]{Corresponding author}\ead{road2you@kisti.re.kr}
\address{Korea Institute of Science and Technology Information \\ 66 Hoegiro, Dongdaemun-gu, Seoul, 02456, Korea}

\begin{abstract}
Academic papers have been the protagonists in disseminating expertise. Naturally, paper citation pattern analysis is an efficient and essential way of investigating the knowledge structure of science and technology. For decades, it has been observed that citation of scientific literature follows a heterogeneous and heavy-tailed distribution, and many of them suggest a power-law distribution, log-normal distribution, and related distributions. However, many studies are limited to small-scale approaches; therefore, it is hard to generalize. To overcome this problem, we investigate 21 years of citation evolution through a systematic analysis of the entire citation history of 42,423,644 scientific literatures published from 1996 to 2016 and contained in SCOPUS. We tested six candidate distributions for the scientific literature in three distinct levels of Scimago Journal \& Country Rank (SJR) classification scheme. First, we observe that the raw number of annual citation acquisitions tends to follow the log-normal distribution for all disciplines, except for the first year of the publication. We also find significant disparity between the yearly acquired citation number among the journals, which suggests that it is essential to remove the citation surplus inherited from the prestige of the journals. Our simple method for separating the citation preference of an individual article from the inherited citation of the journals reveals an unexpected regularity in the normalized annual acquisitions of citations across the entire field of science. Specifically, the normalized annual citation acquisitions have power-law probability distributions with an exponential cut-off of the exponents around 2.3, regardless of its publication and citation year. Our results imply that journal reputation has a substantial long-term impact on the citation. 
\end{abstract}


\end{frontmatter}

%\linenumbers

\section{Introduction}
\noindent Heavy-tailed heterogeneous distributions have been found in many empirical data since their first observation in the economy \cite{Mitzenmacher2004}. Such a distribution is characterized by a long and heavy tail, which is a part of the distribution displaying infrequently occurring events with a large value. Such a highly-skewed distribution seems to be so ubiquitous, from the cyberspace, \textit{i.e.} World Wide Web \cite{Albert1999}, to biological systems, \textit{i.e.} metabolite system of living things \cite{Jeong2000}, being believed as a universal law of nature. Consequently, for almost 20 years, topological studies on large-scale systems have been focused on some interesting features of the scale-free distributions: degree distribution, small-world property, robustness on failure, community structure, core-periphery structure, \textit{etc}. Power law and related distributions have commonly been considered as a holy grail of the complex systems due to their innate properties \cite{Albert2000,Barrat2000}. Scholars occasionally assume that such long-tails imply a power law, which is a quantity x is a result from a probability distribution as follows:
\begin{equation}\label{eq:intro_power-law}
 p(x) \propto x^{-\alpha},
\end{equation}
where $\alpha$ is called an exponent and is a constant parameter that characterizes power-law distribution. Generally, empirical observation follows the exponent in the range $2 < \alpha < 3$, for values larger than a certain minimum $x_{min}$. However, a recent study showed that such a simple power law or complete scale-free distributions are rare \cite{Broido2018}. 

Estimating an appropriate citation distribution is a crucial component of scientometrics to establish unbiased statistical backgrounds for policy and decision makers \cite{Vieira2010, Ruiz2012, Bornmann2017}. Unfortunately, identifying that distribution is challenging, because most citation distributions are characterized as long-tail with rare events, essentially accompanying large fluctuations on observed distributions \cite{Clauset2009}. Although observed data behave like a certain model distribution, it is still hard to deny the possibility of the alternative distributions, e.g. log normals and power laws \cite{Redner1998,Redner2005}. Indeed, scholars proposed several types of model distributions for the citation counts. One possible candidate is power law and its variants \cite{Redner1998,Price1976,Albarran2011,Brzezinski2015}. An exponential and a stretched exponential are also reported \cite{Wallace2009}. In addition, recent studies indicated citation distribution can be explained through the (discretized) lognormal \cite{Thelwall2014, Thelwall2016}. Despite the significant contributions of such endeavours, previous studies mainly focused on the accumulated citation count from the published year to a certain citation window and often neglect the yearly earned citation for the academic literature, e.g. journal articles. There were also several studies investigating dynamics and ageing of citation counts for individual articles \cite{Glanzel2004, Mingers2006, Bouabid2011}; however, these rarely paid attention at the evolution of yearly citation distribution as a whole. For the comprehensive understanding of citation evolution, complementary methodology is necessitated with more in-depth analysis on the increment of citation. 

\begin{figure*}[!ht]
\includegraphics[width=\textwidth]{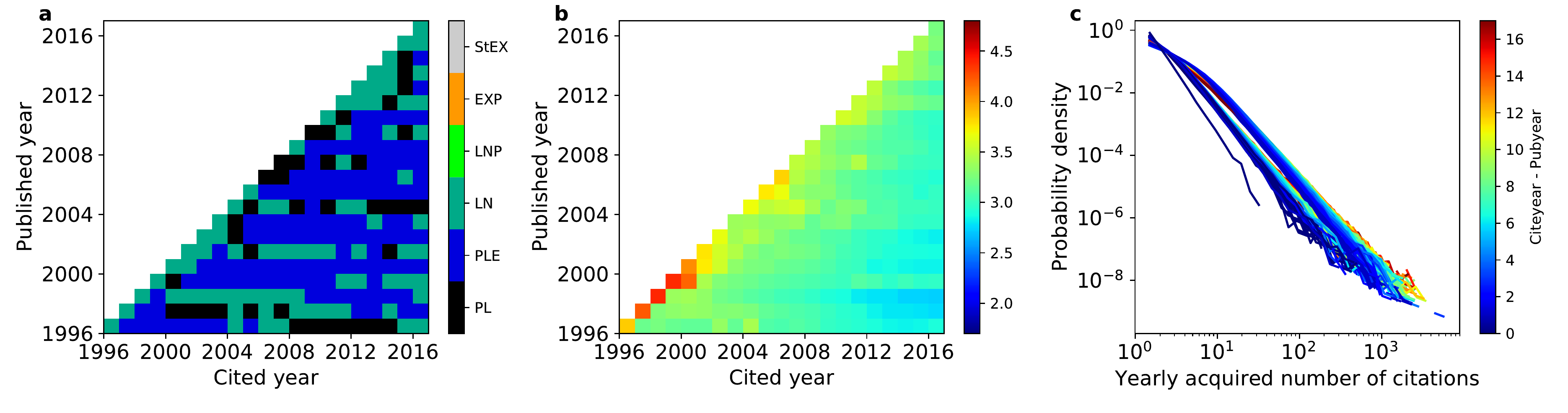}
\caption{Empirical distributions for the number of annual acquired citations, $C_y(a)$. (a) Best fit model distributions obtained through the Maximum Likelihood Ratio methods. For (a), each color denotes the model distributions: simple power law (PL), power law with exponential cut-off (PLE), log normal (LN), log-normal positive (LNP; a case of the log-normal distribution with the value of $\mu > 0$), exponential (EXP), and stretched exponential (StEX). (b) The power-law exponent $\alpha$ for each distribution assuming the power law with an exponential cut-off as the model distribution. (c) Probability density of yearly acquired numbers of citations for the papers. For (c), color denotes the time difference between the year of citations and year of publication: for example, the value of the difference is five if an article published in 2000 cited in 2005. One should note that there also exist the zeroth-year citations, which are cited in the same year as the year of publication.}
\label{fig:bestfit_all}
\end{figure*}

In this study, we perform detailed analysis on the dynamic pattern of citation distributions on the history for 42 million of academic literature published between 1996 and 2016. First, we use the entire citation history to assess the model distribution for yearly acquired citation. Second, we propose journal- and year- normalization method of citation count. Our analysis of articles belonging to distinct disciplines shows that the proper model distribution for the raw counts and normalized counts are different. In this process, we demonstrate that the journal's prestige gives strong impact on the raw citation count so that proper normalization is required to comprehend dynamics of citation evolution. We show that our normalization method can practically remove such a citation surplus owing to the journal's prestige, displayed in the long-term correlation of yearly acquired citation count.

\section{Assessing empirical citation distributions}
\subsection{Data set}\label{sec:data_Set}
\noindent For our analysis on paper metadata, we use the dump of the entire SCOPUS CUSTOM XML DATA for 22 August 2017. This custom data contains the complete copy of data from the SCOPUS website from the very beginning, i.e., January 1996 to August 2017, and includes title, journal, abstract, author information, and citation records in XML format. Each type of document plays different roles in knowledge formation. For example, conference proceedings are a conventional method for presenting new research in the fields of computer science, whereas journal articles are the primary method for many other disciplines. One should note that several disciplines in social science also acknowledge books and reports as essential archives of knowledge. Therefore, to prevent a possible bias towards specific disciplines, we use the entire metadata regardless of the citation type in SCOPUS.

In this data set, there are a total of $42,423,644$ records of academic literature. Each metadata denotes journals and timestamps of the document. This metadata also includes All Science Journal Classification (ASJC) system for each journals \cite{SCOPUSDATA}. This system composite of two-levels hierarchical classifications with 27 subject areas and 334 subject categories, yet some of the subject categories are barely used \cite{Wang2016}. To tackle this issue, we use Scimago Journal \& Country Rank(SJR) consisting of 309 refined subject categories and 27 subject areas from the ASJC scheme \cite{SJR, Gomez2011}. In this study, we take the SJR classification of 2016 regardless of the articles' publication year for the consistency. Journals documented in SJR databases are attributed to least one subject category for ASJC; however, SJR excludes some of the journals with some criteria on the journal quality and journal volume size \cite{SJR, Gomez2011}. Therefore, we also exclude journals not belonging to SJR classification system for the analysis performed on each subject category and area separately. In addition, some journals have multiple IDs due to the altered scope, ISBN or ISSN, publisher, journal title etc. Journal classification may also vary during such changes; thus, we merge the classification information of the journal with multiple IDs, if journals with distinct IDs share an identical journal title. The timestamps of the publications are preferentially extracted from the \texttt{publicationdate} element. It is infrequently replaced by the \texttt{xocs:sort-year} element only if the \texttt{publicationdate} element is empty or incomplete in the metadata. If the timestamp of a certain publication is not between 1996 and 2017, we consider the data as invalid and ignore it. 

\subsection{Best fit model distributions for the annual acquired citation count.}\label{sec:best_fit}
\noindent The power law is characterized with long-tail and rare events, essentially accompanying large fluctuations on observed distributions \cite{Clauset2009}. It is thus hard to infer a suitable distribution from an empirical observation. In fact, even if a power law fits well to the observed data, there is always a possibility of alternative distributions: exponential, log normal, and so on. Moreover, such model distributions might fit better than our primary assumption of power law. Even though there are frequently referred candidates for the heavy-tailed distributions, we choose six candidate models frequently claimed as follows:

\begin{itemize}
\item{Simple power-law distribution
\begin{equation}
p(x) = (\alpha - 1)x^{\alpha-1}_\mathrm{min} x^{-\alpha},
\end{equation}
}
\item{Power-law distribution with an exponential cut-off
\begin{equation}
p(x) = \frac{\lambda^{1-\alpha}}{\Gamma(1-\alpha, \lambda x_\mathrm{min})} x^{-\alpha}e^{-\lambda x},
\end{equation}
}
\item{Exponential distribution
\begin{equation}
p(x) = \lambda e^{\lambda x_\mathrm{min}} e^{-\lambda x},
\end{equation}
}
\item{Stretched exponential distribution
\begin{equation}
p(x) =  \beta \lambda e^{\lambda x_\mathrm{min}^{\beta}} x^{\beta -1} e^{-\lambda x^{\beta}},
\end{equation}
}
\item{Log normal and positive log-normal distributions
\begin{equation}
\begin{aligned}
p(x) = & \sqrt{\frac{2}{\pi \sigma^2}}\left[\mathrm{erfc}\left(\frac{\mathrm{ln} x_\mathrm{min}-\mu}{\sqrt{2}\sigma}\right)\right] ^{-1}\\
       & \qquad \times \frac{1}{x}\mathrm{exp}\left[-\frac{(\mathrm{ln}x - \mu)^2}{2\sigma^2} \right].
\end{aligned}
\end{equation}
}
\end{itemize}

We begin investigating the empirical evidence for best fit model distributions by the Maximum Likelihood Ratio methods \cite{Clauset2009}. We use the yearly acquired number of citations, $C_y(a)$, which is defined as the number of citations of article $a$ obtained in the year $y$. This value implies the level of attention for single academic literature in a particular year, unlike accumulated citations from the published year that reflect the long-term cumulative impact of the academic literature. For the first step, we fitted the empirical distribution of $C_y(a)$ for all six candidate model distributions and each year between 1996 to 2016. For every published and cited years, we scan the parameters including the minimum value of $x$ to find the best fit according to its log-likelihood value, because the citation distribution may not be characterized by a single distribution solely \cite{Redner1998}. This log-likelihood value is also displaying how a particular model distribution is suitable relative to the alternative models. 

\begin{figure*}[!ht]
\includegraphics[width=\textwidth]{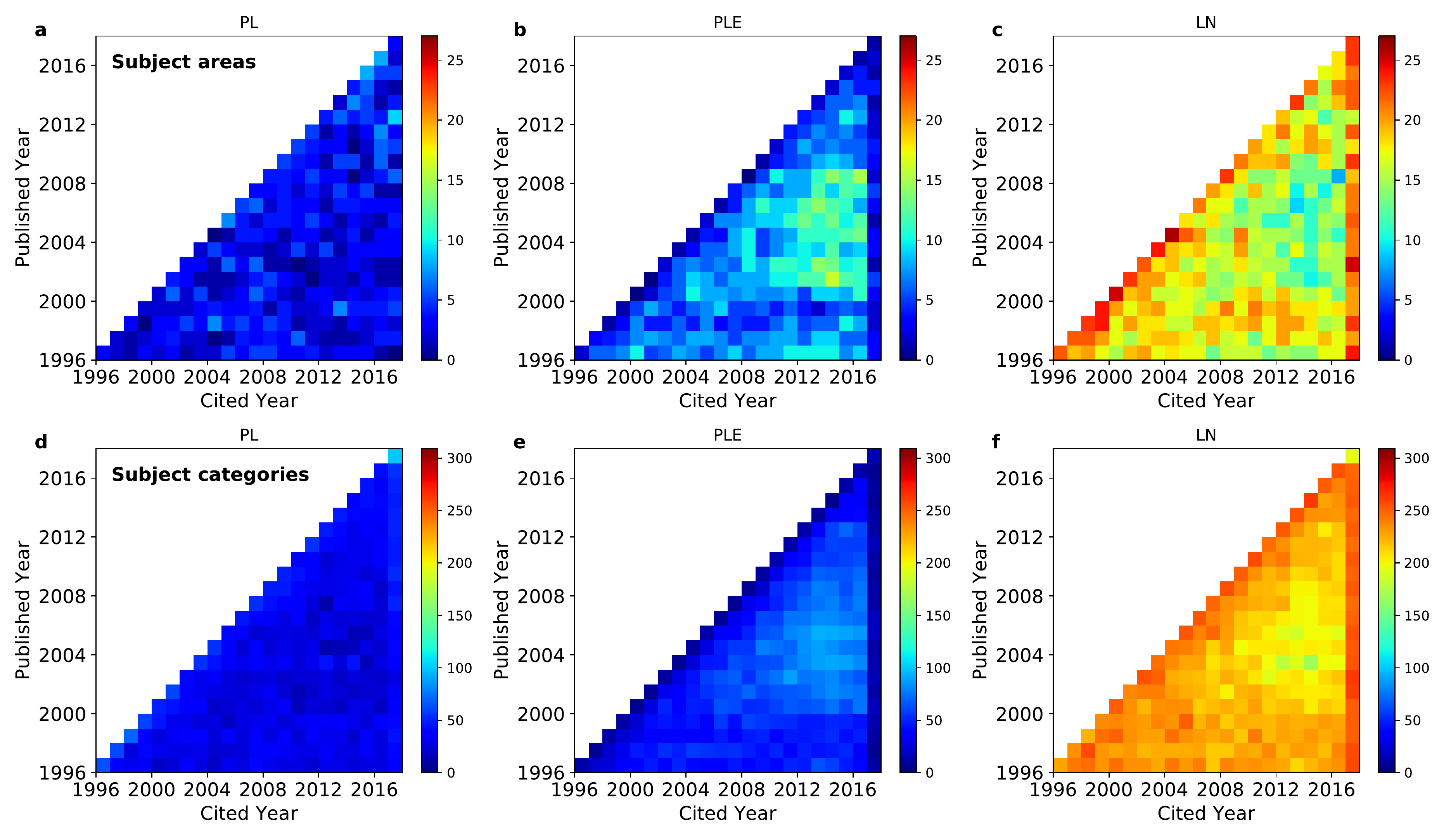}
\caption{Frequency count of best fit model distributions obtained through the Maximum Likelihood Ratio methods for 309 SJR subject categories (a--c) and 27 SJR subject areas (d--f). The fit is obtained for the raw citation count $c_y(a)$. We only display the count for simple power law (PL), the power law with an exponential cutoff (PLE) and log normal(LN), because other three candidates are rarely observed.}
\label{fig:bestfit_asjc_all}
\end{figure*}

For the first step, we apply Maximum Likelihood Ratio methods to the entire publication regardless of its disciplines. Unexpectedly, we observe the mixture of three distributions, instead of the single dominant model (Fig.~\ref{fig:bestfit_all}a). More specifically, we observe the mixture of log normal (LN), power law with an exponential cut-off (PLE), and basic power law (PL) as the best fit of the probability density distribution for the $C_y(a)$. The other three distributions are only observed for negligible numbers. The estimated power-law exponent of $C_y(a)$ ranges extensively from $\sim2.7$ to $\sim4.7$ (Fig.~\ref{fig:bestfit_all}b). This result is also supported by the visual demonstration of probability density showing widespread lines (Fig.~\ref{fig:bestfit_all}c). Therefore, it is hard to conclude any universality of $C_y(a)$ encompassing the different publication years and cited years. 

\begin{figure*}[!ht]
\centering
\includegraphics[width=0.67\textwidth]{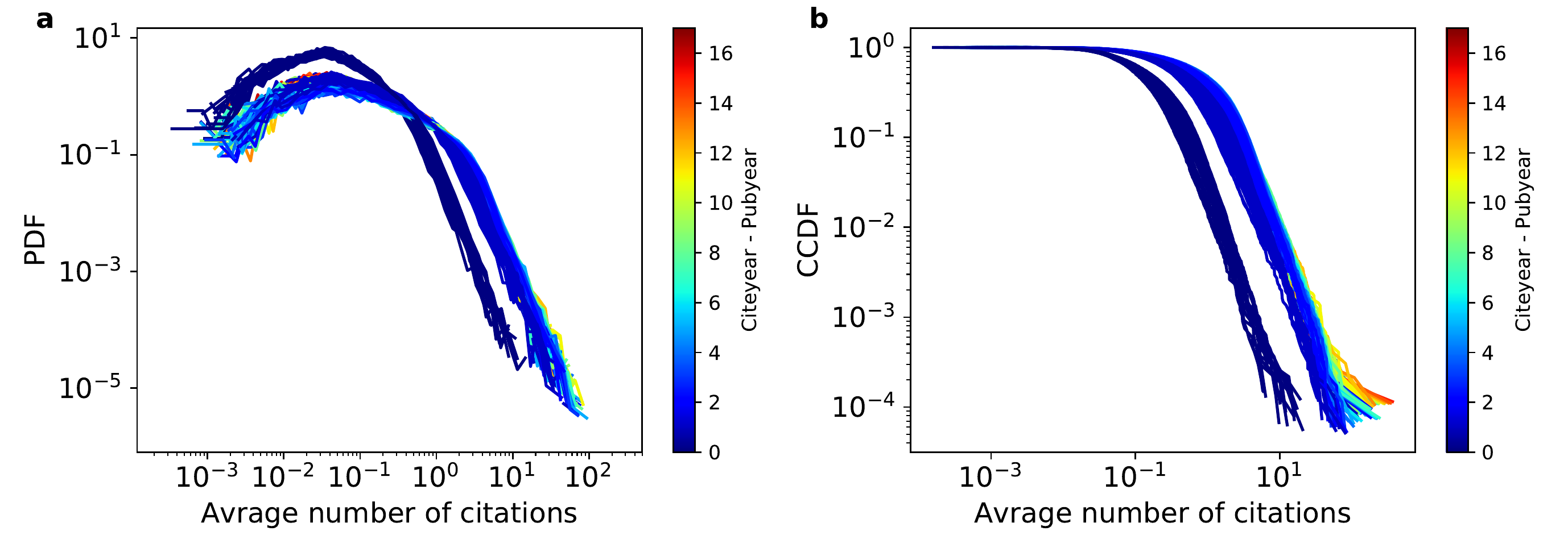}
\caption{Empirical distribution for the average number of yearly acquired citation for each journal: (a) Probability density (PDF) and (b) Complementary cumulative distribution function (CCDF)}
\label{fig:journal_citation_count}
\end{figure*}

One should note that this observation does not completely exclude the possibility of universality. It was reported that one can yield a distribution similar to power law by stacking many log-normal distributions with different means \cite{Stringer2008}. In other words, this mixture of power laws (with or without an exponential cut-off) and log normals may imply the convolution of many log-normal distributions. Indeed, the mean citation count per academic literature varies largely according to its disciplines \cite{Waltman2011}; thus academic citation is the exemplar of the stacking of multiple distributions with different means. Solving the puzzle, we perform a similar analysis of likelihood ratio considering the differences in citation behavior between academic fields. Fig.~\ref{fig:bestfit_asjc_all} shows the count of disciplines that are classified as corresponding model distributions. As suspected, we observe the log normal dominates across the disciplines for every year and both classification levels (see Fig.~\ref{fig:bestfit_asjc_all} c and f). The power law with an exponential cutoff is occasionally detected (see Fig~\ref{fig:bestfit_asjc_all} b and e), whereas basic power law is rarely distinguished (see Fig~\ref{fig:bestfit_asjc_all} a and d). Note that the count of subject categories shows more clear disparity in counts than those of subject areas (compare Fig~\ref{fig:bestfit_asjc_all} a--c and d--f). Stacking more distributions makes it hard to determine the distribution precisely.  

\begin{figure*}[!ht]
\includegraphics[width=\textwidth]{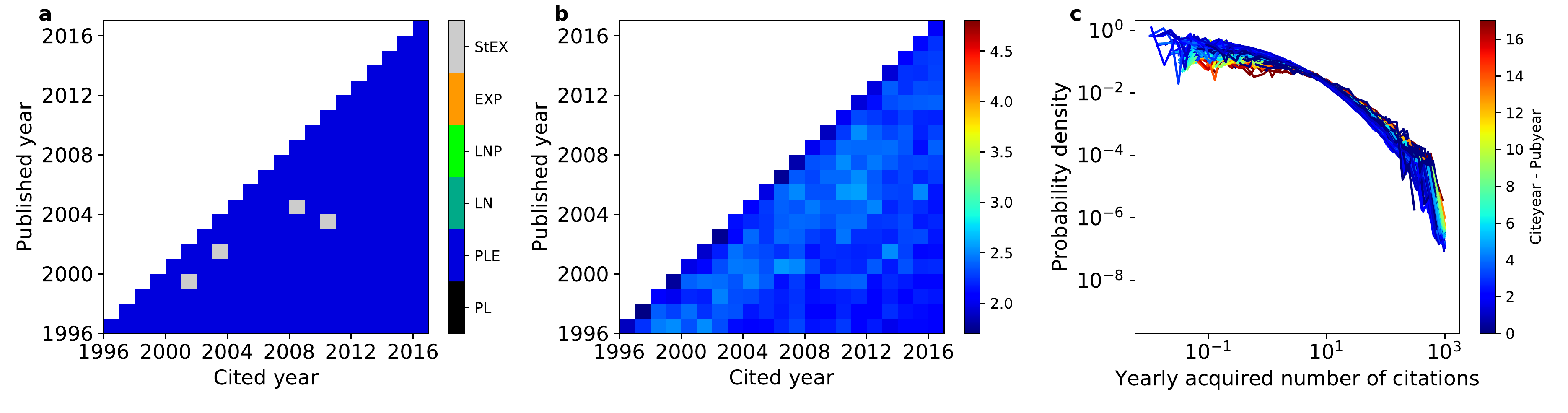}
\caption{Empirical distributions for the normalized count of annual acquired citations $C_y^*(a)$. (a) Best fit model distributions obtained through the Maximum Likelihood Ratio methods. For (a), each color denotes the model distributions: simple power law (PL), power law with an exponential cut-off (PLE), log normal (LN), log-normal positive (LNP; a case of the log-normal distribution with the value of $\mu > 0$), exponential (EXP), and stretched exponential (StEX). (b) Power-law exponent $\alpha$ for each distribution with an assumption of power law with an exponential cut-off as model distribution. (c) Probability density of normalized yearly acquired citations for the papers. For (c), the color denotes the time difference between the year of citations and year of publication: for example, the value of the difference is five if an article published in 2000 cited in 2005. One should note that there also is a zeroth year, which is cited in the same year as the year of publication.}
\label{fig:bestfit_all_norm}
\end{figure*}

\begin{figure*}[!ht]
\includegraphics[width=\textwidth]{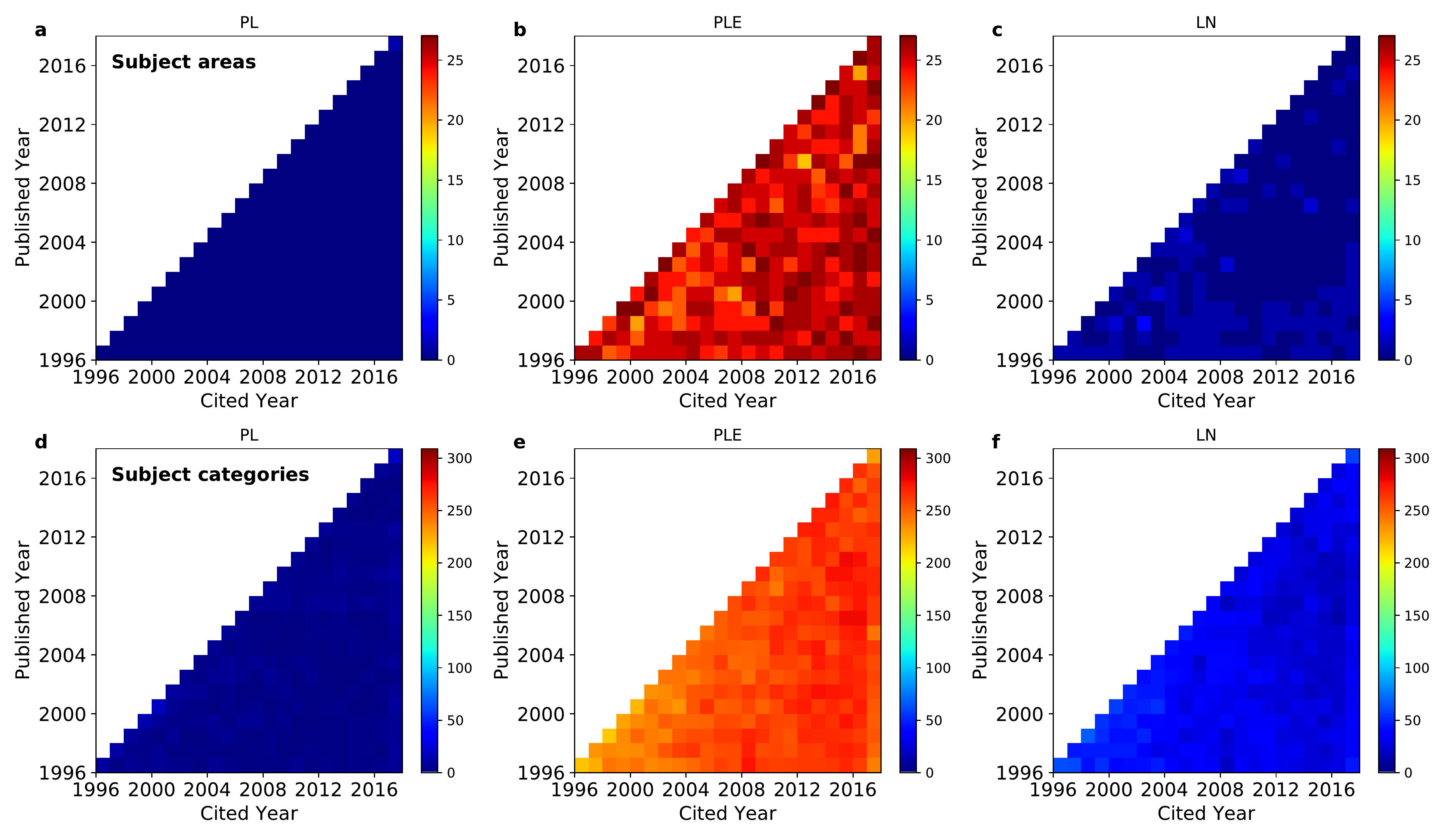}
\caption{Frequency count of best fit model distributions obtained through the Maximum Likelihood Ratio methods for 309 SJR subject categories (a--c) and 27 SJR subject areas (d--f). The fit is obtained for the normalized citation count $C_y^*(a)$. We only display the count for simple power law (PL), the power law with an exponential cut-off (PLE) and log normal (LN), because the other three candidates are barely observed.}
\label{fig:bestfit_asjc_norm}
\end{figure*}

\subsection{Journal- and Time- normalized citation score}\label{sec:journal_mean}
\noindent To proceed with in-depth analysis of citation distribution, we stress the fact that the mean of $C_y(a)$ also varied largely by the journal (see Fig.~\ref{fig:journal_citation_count}), implying the existence of inherited citation due to the prestige of the journal \cite{Lariviere2010, Stegehuis2015}. This background effect itself may make it unfair to directly compare citation counts of articles from different journals and years. Moreover, ageing that consistently reduces the preference of citation was also reported \cite{Eom2011,Hajra2005}. To compensate for such over-representation and ageing effect, we propose the rescaled measures of citation $C^{*}_{y}(a)$ as follows:
\begin{equation}
C^{*}_{y}(a) = \frac{C_y(a)}{\sum_{a \in j(a,y_p)}{C_y(a)}/N\left[j(a,y_p)\right]},
\end{equation}
where $C_y(a)$ is the citation count of article $a$ in the cited year $y$, and $j(a,y_p)$ is the set of articles published in the same journal and published year ($y_p$) of the article $a$. The rescaled citation presents the relative excellence of the academic literature among the most similar publications in terms of the age and journal.

Unlike raw citation, we find that the single distribution dominates with our rescaled citation measure for entire publication and citation year (Fig.~\ref{fig:bestfit_all_norm}a). Across the entire citation and publication years, most plausible distributions are power law with an exponential cut-off, except four citation and publication year pairs (showing stretched exponential; only 1.6\% of the entire pairs). Considering the observation of complex mixture of distributions for the raw citation, such dominance is noteworthy. The estimated power-law exponent of $C^{*}_y(a)$ for those distributions converge around $\sim2.3$ (Fig.~\ref{fig:bestfit_all_norm}b). This finding is also visually supported by the probability density itself, which shows more gathered lines across the years than raw citation $C_y(a)$ (compare Fig.~\ref{fig:bestfit_all_norm}c with Fig.~\ref{fig:bestfit_all}c).

Our analysis on the $C^{*}_y(a)$ considering the differences in citation behavior between academic fields shows the remarkable regularity as well (Fig.~\ref{fig:bestfit_asjc_norm}). Most disciplines (subject area and subject category) exhibit the best-fit distribution as the power law with an exponential cut-off (see Fig.~\ref{fig:bestfit_asjc_all} b and e). The other two distributions are rarely detected (see Fig~\ref{fig:bestfit_asjc_norm} a, c, d, f). One should note that the dominance displayed in the share of best-fit model, log normal for the raw count $C_{y}(a)$ and power law with an exponential cut-off for the $C_{y}^{*}(a)$ respectively, is more prominent in the $C_{y}^{*}(a)$ than in $C_{y}(a)$. Specifically, $92.4\%$ of the total distributions for the $C^{*}_{y}(a)$ are observed as power law with an exponential cut-off among the subject categories (6\,313 distributions out of 6\,831 distributions), whereas only 63.7\% of the distributions showing log-normal behavior for the $C_y(a)$ (4\,355 distributions out of 6\,831 distributions). The experiment across subject categories shows a bit weak, but similar result with the case of subject areas: only 73.0\% of the distributions of $C_y(a)$ are found to be log normal (57\,040 out of 78\,177 distributions), yet 83.4\% of the $C^{*}_y(a)$ distributions are power law with an exponential cutoff. 

\begin{figure*}[!ht]
\centering
\includegraphics[width=0.67\textwidth]{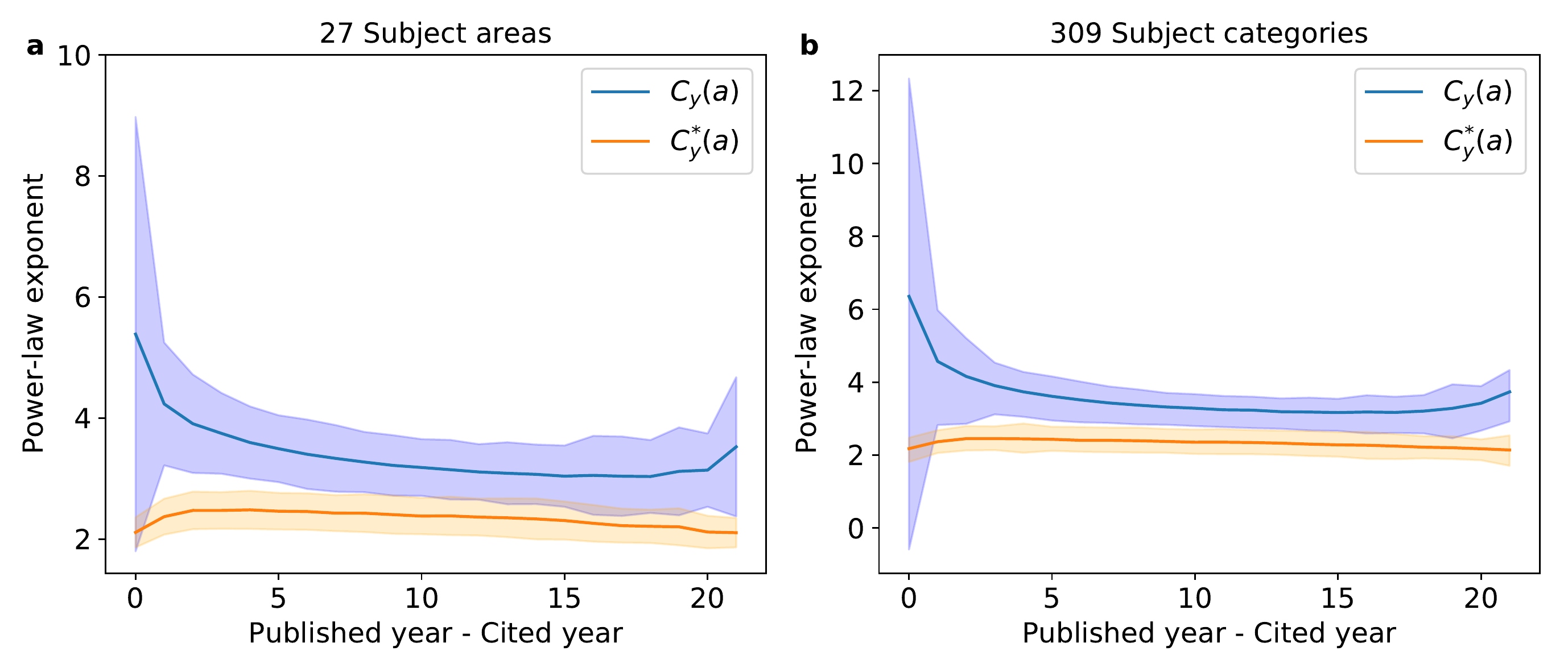}
\caption{Power-law exponents of raw citation count $C_y(a)$ and rescaled citation count $C^{*}_y(a)$. The exponent is averaged over 27 subject areas (a) and 309 subject categories (b) for every citation time difference between the published year and cited year. The shaded area corresponds to standard deviation over different disciplines and published years.}
\label{fig:avg_exponent}
\end{figure*}

Although the raw citation $C_y(a)$ is hardly considered as power law (with an exponential cut-off) according to aforementioned observations, it is still worthwhile to measure the power-law exponent because it can be used as a proxy of the heterogeneity for probability distributions \cite{Hu2008}. The above estimation of exponent suggests that raw citation distribution is becoming more heterogeneous as the time passed from the publication (Figs~\ref{fig:bestfit_all}c). The exponent of normalized measure $C^{*}_y(a)$ changes relatively smaller than those of raw citation distribution (Fig.~\ref{fig:bestfit_all_norm}c). A logical step forward is to search for the fluctuation degree regarding the disciplines. In Fig.~\ref{fig:avg_exponent}, we discover that the exponent of rescaled citations is not only less fluctuated with the time difference between the year of citation and publication but also more stable with the disciplines than the raw citation measure $C_y(a)$. Therefore, the rescaled measures are nearly free from the journal and year effect, making it possible to compare a significant amount of scientific literature from different years and journals into the same place for analysis away from the inherited impact of the journals. 

\subsection{Memory effect of $C_{y}(a)$ and $C^{*}_{y}(a)$}\label{sec:memory_effect}
\noindent Identifying the emerging concept of science and technology is obviously the desired goal, both for the researchers and policymakers. Scanning \textit{highly cited papers} is a common tool for the sensing the emerging or breakthrough concept \cite{Oppenheim1978,Aksnes2003,Schneider2017}; however, citation behaviors differ between disciplines \cite{Waltman2011}. Thus, the definition and interpretation of a concept \textit{highly cited paper} is complicated. Moreover, the existence of the ageing effect raises the degree of difficulty for comparing the citations from the different years \cite{Glanzel2004,Mingers2006, Bouabid2011,Eom2011,Hajra2005}. The proposed rescaled measure $C^{*}_y(a)$ successfully reduce the influence of the discipline and ageing (see Fig.~\ref{fig:bestfit_asjc_norm} and Fig.~\ref{fig:avg_exponent}, respectively). In short, we successfully minimize the discussed fluctuations with a simple normalization. 

\begin{figure*}[!ht]
\includegraphics[width=\textwidth]{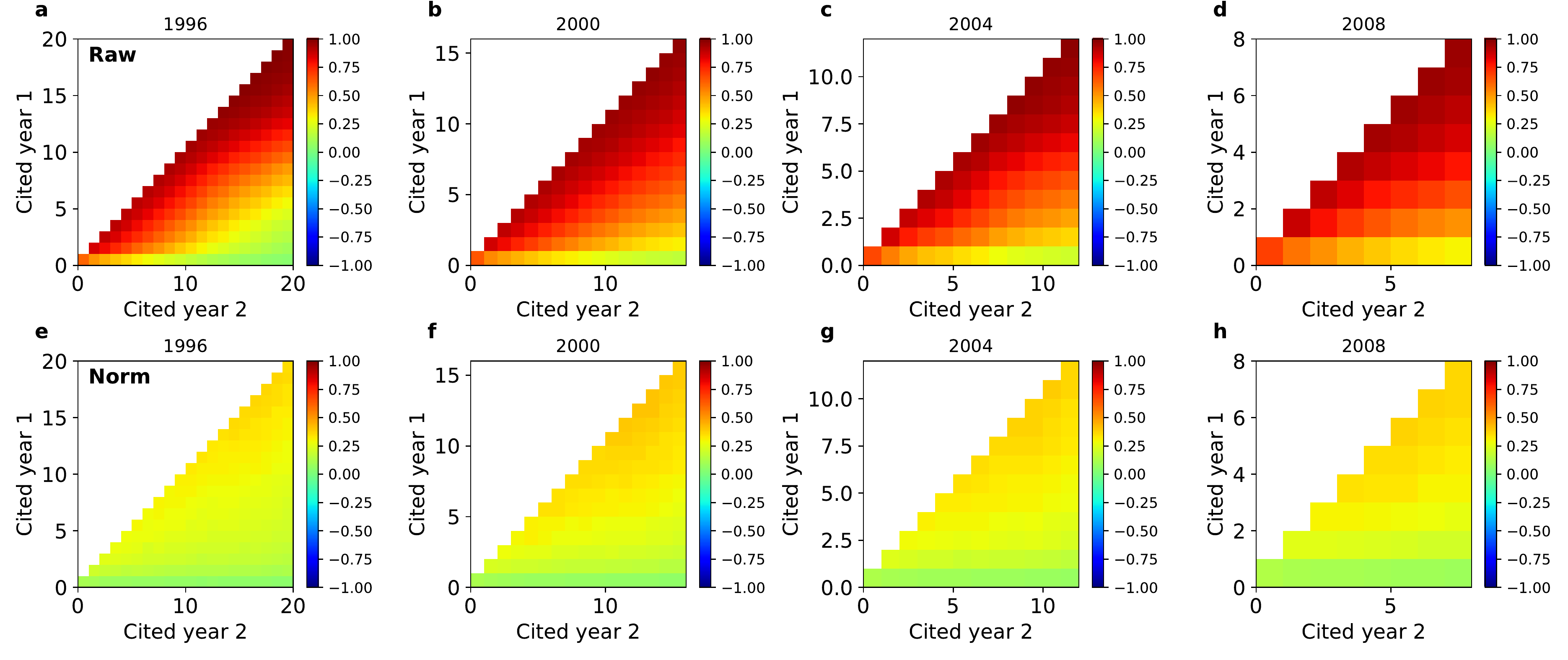}
\caption{The Pearson correlation coefficient between the two lists of the citation count, acquired by a specific scientific literature between two given cited years from the published year on each axis (a)--(d) for raw citation $C_y(a)$ and (e)--(h) for normalized citation $C^{*}_y(a)$. Texts at the top of each panel denote the published year. Both axes indicate the year of citation from the publication, e.g. 10 indicates 2006 for the articles published in 1996. One should note that the citation distribution is heavy-tailed and zero-inflated, so the influence of non-cited article is large for the correlation value. Thus, for this figure, we only considered articles that are cited at least once in any of two cited years, to avoid an overestimated impact from non-cited articles.}
\label{fig:temporal_memory}
\end{figure*}

One last remained property of citation behavior is the rich-get-richer phenomena \cite{Borner2004, Wang2014}, possibly influenced by the fame of the journals belonging to, as papers published in headliner journals easily get the early citation. To probe this, we use Pearson's correlation between two different cited years, from the lists of the papers published in a certain year. As we expected, raw citation shows the strong correlation between two citation years (Fig.~\ref{fig:temporal_memory}a--d). The influence of early citation lasts more than a decade; meanwhile citations of more than decade after publication still have strong correlation with the citation in later periods. (Fig.~\ref{fig:temporal_memory}a--d). Considering that the later citation has influenced by the earlier ones, the impact of early citation cannot be neglected. On the contrary, the sequence of rescaled citation count $C^{*}_y(a)$ shows insignificant correlations across the cited years (Fig.~\ref{fig:temporal_memory}a--d). Additionally, the impact of early citation is almost zero, so that be seen as a no effect of early citation (see the correlations of Cited year 2 $\leq 5$ in Fig.~\ref{fig:temporal_memory}). Gathering up the threads, we find that the influence from the initial citation lasts long, but it may be neutralized by our normalization method.

\section{Discussion}
\noindent In this study, we investigated the structure of academic citation through a massive history of citation metadata over the past two decades magnifying the influence of the journals' prestige. Our finding suggests that citation evolution is not solely affected by the influence of paper itself but by the overall influence of the attributes: discipline, ageing, and early citation due to the journals' prestige. We also supplement the evidences for the influence of the journals on the citations that would enhance the merit of previous studies \cite{Stegehuis2015,Didegah2013,Tahamtan2016}. We believe that extending our analysis into various agents, e.g. impact of countries, authors, institutes, and disciplines, is necessary to understand the ecosystem of science and technology deeper, yet we leave the tasks for further study.

Our approach also has notable implications for policy-making, especially when collaborated with other elaborate methodologies \cite{Shibata2008}. While evaluations for scientific investments are conventionally based on research achievements, e.g., the number of citations, publications, and the reputations from colleagues, citation boosting by the fame of authors, journals, countries, and other agents are easily overlooked. Even though we investigated the influence from the journals only, our results imply that the influence of \textit{halo effect}, \textit{i.e.} citation attributed by the environment, lasts for a long time. Such an impact also may result in less accuracy of evaluations; thus, a comprehensive understanding of the factors is demanded to set an unbiased and fair standard. Beyond the impact on the citation that we harnessed in this study, a myriad of online resources responding to science and technology also can be rescaled from the similar spirit because altmetric scores can also be influenced by the fame of early spreaders. Going one step forward, we would like to emphasize that data-driven analysis should be accompanied by proper normalization and aptly integrated with the contents-oriented and qualitative perspectives of approaches that span the entire progress of knowledge accumulation \cite{Waltman2016, Leydesdorff2016a}. If the task is accomplished, the synergy will bring for the application of citation analysis. Finally, we also hope that our approach sheds light on the unbiased understanding of citation dynamics of the science and technology in the future.

\section{Author's contribution}
Jinhyuk Yun: Conceived and designed the analysis; Collected the data; Performed the analysis; Wrote the paper.
Sejung Ahn: Conceived and designed the analysis; Wrote the paper.
June Young Lee: Conceived and designed the analysis; Collected the data; Wrote the paper.

\section{Acknowledgement}
This work was supported by the National Research Foundation of Korea Grant funded by the Korean Government through Grant No. NRF-2017R1E1A1A03070975 (J.Y.; S.A.) and the Korea Institute of Science and Technology Information. The funders had no role in study design, data collection and analysis, decision to publish, or preparation of the manuscript.

%\section*{References}
\bibliography{power-law-of-articles}
\end{CJK}
\end{document}